\documentclass{aa}  

\usepackage{graphicx}
\usepackage{txfonts}
\begin{document}

   \title{Accretion of eroding pebbles and planetesimals in planetary envelopes}
   \titlerunning{Accretion of eroding pebbles and planetesimals in planetary envelopes}

   \author{Tunahan Demirci
          \and
          Gerhard Wurm}

   \institute{University of Duisburg-Essen, Faculty of Physics, Lotharstr. 1, 47057 Duisburg, Germany\\
              \email{tunahan.demirci@uni-due.de, gerhard.wurm@uni-due.de}
             }

   \date{Received 18 June 2020 / Accepted 2 July 2020}

        \abstract{Wind erosion is a destructive mechanism that completely dissolves a weakly bound object like a planetesimal into its constituent particles, if the  velocity relative to the ambient gas and the local gas pressure are sufficiently high. In numerical simulations we study the influence of such wind erosion on pebble and planetesimal accretion by a planetary body up to $10 R_\mathrm{Earth}$. Due to the rapid size reduction of an in-falling small body, the accretion outcome changes significantly. Erosion leads to a strong decrease in the accretion efficiency below a threshold size of the small body on the order of 10 m. This slows down pebble accretion significantly for a given size distribution of small bodies. The threshold radius of the small body increases with increasing planet radius and decreases with increasing semi-major axis. Within the parameters studied, an additional planetary atmosphere (up to 1 bar) is of minor importance.} 


   \maketitle
%

\section{Introduction}
\label{sec:introduction}

In current scenarios of planet formation, a small planetary body grows into a full-sized planet by accreting smaller bodies that might vary in size from pebbles to planetesimals \citep{Ormel2010, Lambrechts2012, Johansen2017, Bitsch2018, Bitsch2019, Liu2019, Bruegger2020}. The basic idea is that a planetary body encounters smaller pebbles and planetesimals on its orbit, which are then gravitationally attracted by the growing planet and are added to its solid mass. Depending on the size and the local gas pressure, the small bodies experience gas drag. Very small grains or dust couple to the gas perfectly. They follow streamlines around the planet and are not accreted. Pebble-sized objects are efficiently slowed down and eventually get the chance to settle onto the planet. Planetesimals are only slightly affected by the gas drag, so accretion efficiency is reduced again, although a single planetesimal homing in on the planet can deliver more mass. Pebble and planetesimal accretion are two terms often encountered in this context in the literature but they are only two extremes of the same story, each focusing on a specific size scale. Hybrid accretion scenarios with both size scales are also viable options \citep{Alibert2018, Bruegger2020}. We will also consider accretion of different-sized bodies here.\\

Due to gas drag, accretion is strongly dependent on the size of the incoming body. Therefore, a change in size during the accretion process will change the accretion outcome significantly. It has sometimes been considered that planetesimals on supersonic trajectories become ablated \citep{Valletta2019}. This changes their size especially if they break up at a certain point, though this is considered more in the context of providing the atmosphere with heavier elements \citep{Venturini2020}.\\

On a more elemental level, for weakly bound bodies moving through a protoplanetary disk, wind erosion occurs long before ablation has a chance to take over. All primordial bodies smaller than a large planetesimal are thought to consist of dust and therefore are weakly bound.
First generation small bodies are thought to form from small millimeter-sized dust aggregates \citep{Kruss2016}, which are only put together loosely as they are concentrated by various drag instabilities or the streaming instability \citep{Youdin2005, Johansen2015, Simon2016, Yang2017, Schreiber2018, Squire2018, Schneider2019}. This results in low tensile strength \citep{Skorov2012}.

The stability of small weak bodies against wind erosion has been studied experimentally by \citet{Demirci2019} and \citet{Demirci2020}. They found that small bodies on a regular circular orbit in the inner protoplanetary disk are already affected by erosion up to complete destruction. \citet{Schaffer2020} also recently used numerical simulations to study the conditions under which planetesimals are eroded by gas drag on eccentric orbits. 
To put this in context here, the head wind that a body encounters on a circular orbit is about $50\,\mathrm{m}\,\mathrm{s}^{-1}$, but if it is accelerated by a planet's gravity, like in the pebble accretion scenario, the head wind and therefore the surface shear stress increases. This makes pebbles and planetesimals, which were stable on their orbit, prone to wind erosion in the planetary envelopes.

In this work, we simulate the influence of wind erosion on pebble and planetesimal accretion.

\section{Numerical method}

\begin{figure}
        \centering
        \includegraphics[width=\columnwidth]{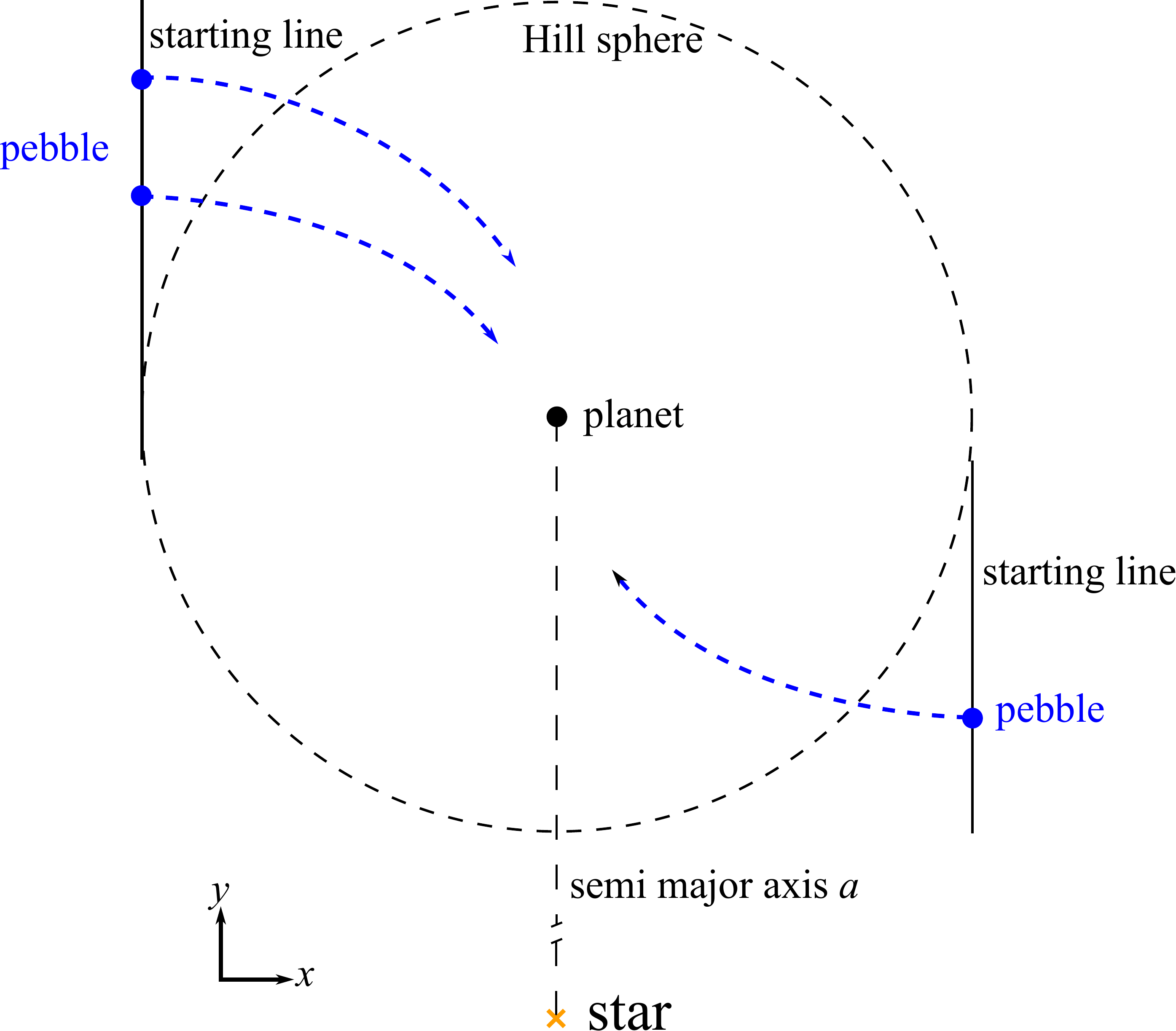}
        \caption{Simulation geometry. The planet is fixed to the center of the rotating coordinate system. The initial positions of the pebbles are distributed on the starting lines. At the beginning of the simulation the pebbles have the same velocity as the gas in the same location. The paths of the pebbles evolve in time according to the equation of motion (Eq. \ref{eq:motionequation}).}
        \label{fig:setup}
\end{figure}

In this section we introduce the simulation geometry and the equations that are considered for the numerical simulations of different pebble accretion scenarios. Figure \ref{fig:setup} shows the rotating coordinate system, where the planet is fixed to the center. The planet has a radial distance $a$ to the star in a minimum mass solar nebula \citep{Hayashi1981}. The gas density at the planet's location is
\begin{equation}
\rho_\mathrm{PPD}=\rho_0 \left( \frac{a}{1\,\mathrm{AU}} \right)^{-\frac{11}{4}},
\end{equation}
with $\rho_0=1.4 \times 10^{-6}\,\mathrm{kg}\,\mathrm{m}^{-3}$ being the gas density at $1\,\mathrm{AU}$. If the planet has an atmosphere, we take the total gas density at a radial distance $r$ to the planet as
\begin{equation}
\label{eq:totalgaspressure}
\rho_\mathrm{total}(r,a)=\rho_\mathrm{PPD}(a)+\rho_\mathrm{atmosphere}(r,a),
\end{equation}
with 
\begin{equation}
\rho_\mathrm{atmosphere}(r,a)=\rho_\mathrm{surface} \cdot \exp \left(  -\frac{r-R_\mathrm{planet}}{R_\mathrm{s} T(a)} g_\mathrm{planet} \right)
\end{equation}
following the barometric formula. It depends on the gas density $\rho_\mathrm{surface}$ and the gravitational $g_\mathrm{planet}$ acceleration at the planet's surface, and the local temperature $T(a) = 280\,\mathrm{K}\left( \frac{a}{1\,\mathrm{AU}} \right)^{-\frac{1}{2}}$ \citep{Hayashi1981}. The specific gas constant of $\mathrm{H}_2$ is described by $R_\mathrm{s}$.

At the radial distance $a$ the Kepler frequency $\Omega_0$ and Hill radius $r_\mathrm{Hill}$ of the planet are
\begin{equation}
\Omega_0(a)=\left(\frac{G m_\mathrm{star}}{a^3}\right)^{\frac{1}{2}}\mathrm{and}
\end{equation}
\begin{equation}
r_\mathrm{Hill}(a)=a \left(\frac{m_\mathrm{planet}}{3 m_\mathrm{star}} \right)^{\frac{1}{3}},
\end{equation}
with the gravitational constant $G$, the stellar mass $m_\mathrm{star}$, and the planet mass $m_\mathrm{planet}$.

The relative velocity between objects on circular Kepler orbits and the gas is only weakly dependent on the semi-major axis $a$. For the minimum mass solar nebula, it is on the order of $u_\mathrm{rel}=50\,\mathrm{m}\,\mathrm{s}^{-1}$ \citep{Weidenschilling1977}. In the rotating coordinate system the relative gas velocity is expressed as $\mathbf{u}_\mathrm{rel}+\frac{3}{2}y \Omega_0 \mathbf{e_x}$, caused by the increasing Kepler frequency $\Omega_0$ for decreasing $y$ (and vice versa) \citep{Ormel2010}. To take a non-moving planetary atmosphere into account, we weigh its influence on the total relative gas velocity in this coordinate system with the local gas density  
\begin{equation}
\label{eq:gasvelocity}
\mathbf{u}_\mathrm{gas}(r,a)=\frac{\rho_\mathrm{PPD}(a)}{\rho_\mathrm{PPD}(a)+\rho_\mathrm{atmosphere}(r,a)}\left(\mathbf{u}_\mathrm{rel}+\frac{3}{2}y \Omega_0 \mathbf{e_x}  \right).
\end{equation}
Far from the planet the velocity is $\mathbf{u}_\mathrm{rel}+\frac{3}{2}y \Omega_0 \mathbf{e_x}$, but near to the planetary surface, where the atmospheric gas density $\rho_\mathrm{atmosphere}(r,a)$ becomes significant, the relative gas velocity converges to zero for dense atmospheres.\\

The equation of motion in a rotating system, where the planet is fixed to the center of the coordinate system, is \citep{Ormel2010}
\begin{equation}
\label{eq:motionequation}
\ddot{\mathbf{r}}=\frac{\mathbf{u}_\mathrm{gas}(r,a)-\dot{\mathbf{r}}}{t_\mathrm{c}(r,a)}-G m_\mathrm{planet} \frac{\mathbf{r}}{r^3}-2 \Omega_0 \times \dot{\mathbf{r}}+ \Omega_0^2 \mathbf{r}.
\end{equation}
The equation of motion for the pebbles includes the gas drag, the gravitational interaction with the planet, and the Coriolis and centrifugal force caused by the rotating system. The coupling time of the pebble to the gas is $t_\mathrm{c}$  and is described as
\begin{equation}
\label{eq:coulingtime}
t_\mathrm{c}=\left\{\begin{array}{ll} f_\mathrm{C} \cdot \frac{2}{9} \cdot \frac{\rho_\mathrm{pebble} d_\mathrm{pebble}^2}{\eta} &\mathrm{ for }~\mathrm{Re}<0.1,\\
\\
\frac{f_\mathrm{C}}{C_\mathrm{d}} \cdot \frac{4}{3} \cdot  \frac{\rho_\mathrm{pebble}}{\rho_\mathrm{total}} \frac{d_\mathrm{pebble}}{\Delta v} &\mathrm{ for }~\mathrm{Re}\geq 0.1.\\
\end{array}
\right.
\end{equation}
Here, we distinguish between two cases, the Stokes regime $\mathrm{Re}<0.1$ and the quadratic regime $\mathrm{Re} \geq 0.1.$ For small Reynolds numbers, $\mathrm{Re}=\rho \Delta v d_\mathrm{pebble}/\eta$, with $\eta$ being the dynamic viscosity of the gas; the drag force is linearly dependent on the relative velocity $\Delta v$  between the pebble and gas, and the coupling time can be expressed independently from $\Delta v$. This changes for larger Reynolds numbers. Here, the drag force depends on $\Delta v^2$. The pebble diameter is
$d_\mathrm{pebble}$  and $C_\mathrm{d}$ is the drag coefficient for a sphere and is taken from \citet{Brown2003},
\begin{equation}
C_\mathrm{d}(\mathrm{Re})=\frac{24}{\mathrm{Re}}\left(1+0.15\mathrm{Re}^{0.681}\right)+\frac{0.407}{1+8700\mathrm{Re}^{-1}},
\end{equation}
and is valid for Reynolds numbers $\mathrm{Re}< 2\times 10^5$. To take molecular flow effects into account, both cases in Eq. \ref{eq:coulingtime} are scaled with the Cunningham correction \citep{Davies1945}
\begin{equation}
f_\mathrm{C}(\mathrm{Kn})=1+2 \mathrm{Kn} \left[1.257 + 0.4 \exp\left(- \frac{0.55}{\mathrm{Kn}}  \right)  \right].
\end{equation}
The Cunningham correction is a function of the Knudsen number of the pebbles 
\begin{equation}
\mathrm{Kn}= \frac{\lambda}{d_\mathrm{pebble}} =\frac{k_\mathrm{B}}{\sqrt{2} \pi d_\mathrm{mol}^2 R_\mathrm{s} \rho_\mathrm{total} d_\mathrm{pebble}},
\end{equation}
which describes the ratio of the mean free path $\lambda$ of the gas molecules and the pebble diameter $d_\mathrm{pebble}$. Here, $k_\mathrm{B}$ is the Boltzmann constant and $d_\mathrm{mol}$ the molecular diameter. For large Knudsen numbers the coupling time describes the case of the Epstein regime.\\

To study the size evolution of small bodies, we numerically integrated the equation of motion (Eq. \ref{eq:motionequation}). A short code for this was programmed dedicated to the problem. We used the fourth-order Runge-Kutta integration method with an adaptive time step, which ensures a velocity resolution of $v_\mathrm{res}=0.1\,\mathrm{m}\,\mathrm{s}^{-1}$. However, the time step never exceeds $0.3t_\mathrm{c}$. At the beginning of the simulation runs the pebbles are distributed on the starting lines (see Fig. \ref{fig:setup}). These starting lines are located at $(+r_\mathrm{Hill},y,0)$ for $y \in [-b_0~ r_\mathrm{Hill},b~r_\mathrm{Hill}]$ and $(- b~r_\mathrm{Hill},y,0)$ for $y \in [- b ~r_\mathrm{Hill},- b_0~ r_\mathrm{Hill})$, with $b_0=2 u_\mathrm{rel}/\left(3 r_\mathrm{Hill} \Omega_0\right)$ and $b$ as constant, which must be selected as a sufficiently large quantity that the entire accretion cross-section $\sigma$ is considered within the simulation parameters. At the beginning of the simulation the pebbles have the same velocity as the surrounding gas (see Eq. \ref{eq:gasvelocity}).\\

We are interested in the influence of wind erosion on pebble accretion. Therefore, the considered pebbles in our simulation are porous clusters consisting of particles of a diameter of $d_\mathrm{particle}=10^{-3}\,\mathrm{m}$ (bouncing barrier size) \citep{Kruss2016,Kruss2017,Demirci2017}. For each time step in the simulation, the shear stress $\tau_\mathrm{wall}$, which is acting on the pebble, is compared with the shear stress $\tau_\mathrm{erosion}$ for wind erosion. The wall shear stress $\tau_\mathrm{wall}=\eta u_\mathrm{gas}/\delta$ can be determined by approximating the boundary layer thickness $\delta=5\sqrt{\eta l/\Delta v}$ around a sphere and is expressed as \citep{Schlichting2006}
\begin{equation}
\tau_\mathrm{wall}=\frac{1}{5}\left(\frac{2 \rho~ \eta ~\Delta v^3}{d_\mathrm{pebble}} \right)^{\frac{1}{2}}.
\end{equation}
The shear stress $\tau_\mathrm{erosion}$, which is needed to erode a protoplanetary surface, was determined experimentally for a wide range of gas pressure and gravitational acceleration by \citet{Demirci2019} and \citet{Demirci2020}. It is
\begin{equation}
\tau_\mathrm{erosion}=\alpha f_\mathrm{C}(\beta^{-1} \mathrm{Kn})\left( \frac{\gamma_\mathrm{eff}}{d_\mathrm{particle}} + \frac{1}{9} \rho_\mathrm{pebble} g_\mathrm{pebble} d_\mathrm{particle} \right),
\end{equation}
where the first term describes the cohesion and the second term describes the gravitational acceleration. The gravity term only becomes  important for larger objects, like planetesimals. The effective surface energy $\gamma_\mathrm{eff}$ \citep{Johnson1971} of the constituent particles is assumed to be about $10^{-4}\,\mathrm{N}\,\mathrm{m}^{-1}$ and $\beta=0.67$ is an empirical scaling factor for the Knudsen number in the Cunningham correction for $d_\mathrm{particle}$ \citep{Demirci2019,Demirci2020}. If the wall shear stress exceeds the erosion shear stress ($\tau_\mathrm{wall}>\tau_\mathrm{erosion}$), the radius of the pebble will be decreased with a certain erosion rate $\epsilon=\Delta R_\mathrm{pebble}/\Delta t$. We take $\epsilon=5\times 10^{-4}\,\mathrm{m}\,\mathrm{s}^{-1}$, which was observed experimentally near the threshold shear stress by \citet{Demirci2019}. A pebble could be eroded by the gas drag - if the conditions are fulfilled - to a size down to that of one particle $d_\mathrm{particle}=10^{-3}\,\mathrm{m}$.

\section{Results}

\begin{table}
        \caption{Simulations parameters. For each simulation run the pebble size is varied between $10^{-3}$ and $10^{4}\,\mathrm{m}$ and the range for the initial $y$-position is kept large enough so that all accretion events are considered in the simulations.}
        \label{tab:simparameter}
        \centering
        \begin{tabular}{ccc}
                \hline \hline
                Planet radius &  Semi-major axis& Gas pressure at\\
                &  & planetary surface\\
                $R_\mathrm{planet}~[R_\mathrm{Earth}]$ & $a~[\rm AU]$ & $p_\mathrm{surface}~[\mathrm{Pa}]$\\
                \hline \hline
                $0.1$ & $0.1, 0.2, 0.3, 0.5,$ & $0$\\
                & $1,2,3$ & \\
                \hline
                $0.2$ & $0.1, 0.2, 0.3, 0.5,$ & $0$\\
                & $1,2,3$ & \\
                \hline
                $0.3$ & $0.1, 0.2, 0.3, 0.5,$ & $0$\\
                & $1,2,3$ & \\
                \hline
                $0.5$ & $0.1, 0.2, 0.3, 0.5,$ & $0$\\
                & $1,2,3$ & \\
                \hline
                $1$ & $0.1, 0.2, 0.3, 0.5,$ & $0$\\
                & $1,2,3$ & \\
                \hline
                $2$ & $0.1, 0.2, 0.3, 0.5,$ & $0$\\
                & $1,2,3$ & \\
                \hline
                $3$ & $0.1, 0.2, 0.3, 0.5,$ & $0$\\
                & $1,2,3$ & \\
                \hline
                $5$ & $0.1, 0.2, 0.3, 0.5,$ & $0$\\
                & $1,2,3$ & \\
                \hline
                $10$ & $0.1, 0.2, 0.3, 0.5,$ & $0$\\
                & $1,2,3$ & \\
                \hline
                $1$ & $0.1, 0.2, 0.3, 0.5,$ & $10^3$\\
                & $1,2,3$ & \\
                \hline
                $1$ & $0.1, 0.2, 0.3, 0.5,$ & $10^5$\\
                & $1,2,3$ & \\
                \hline \hline
        \end{tabular}
\end{table}

\begin{figure}
        \centering
        \includegraphics[width=\columnwidth]{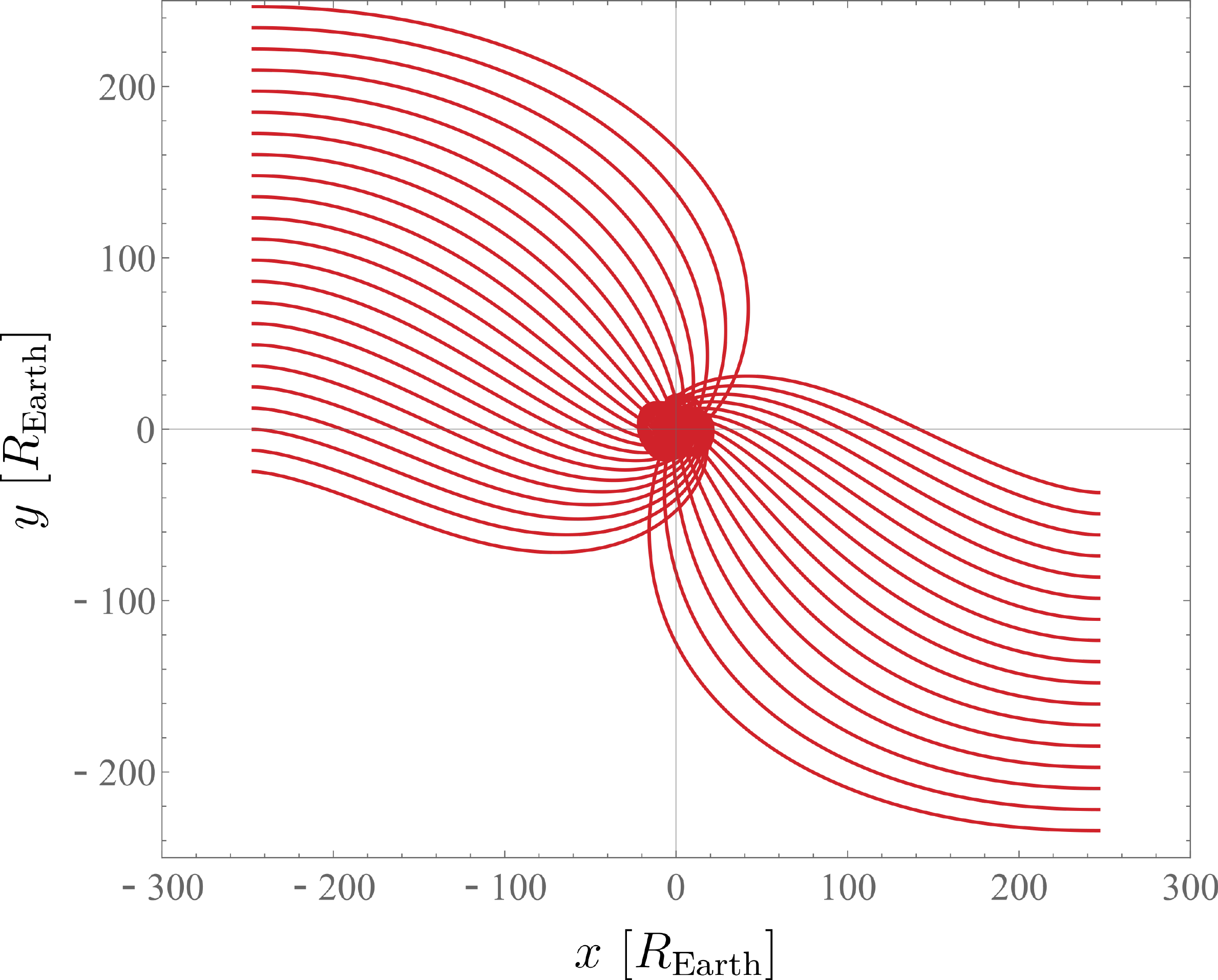}
        \includegraphics[width=\columnwidth]{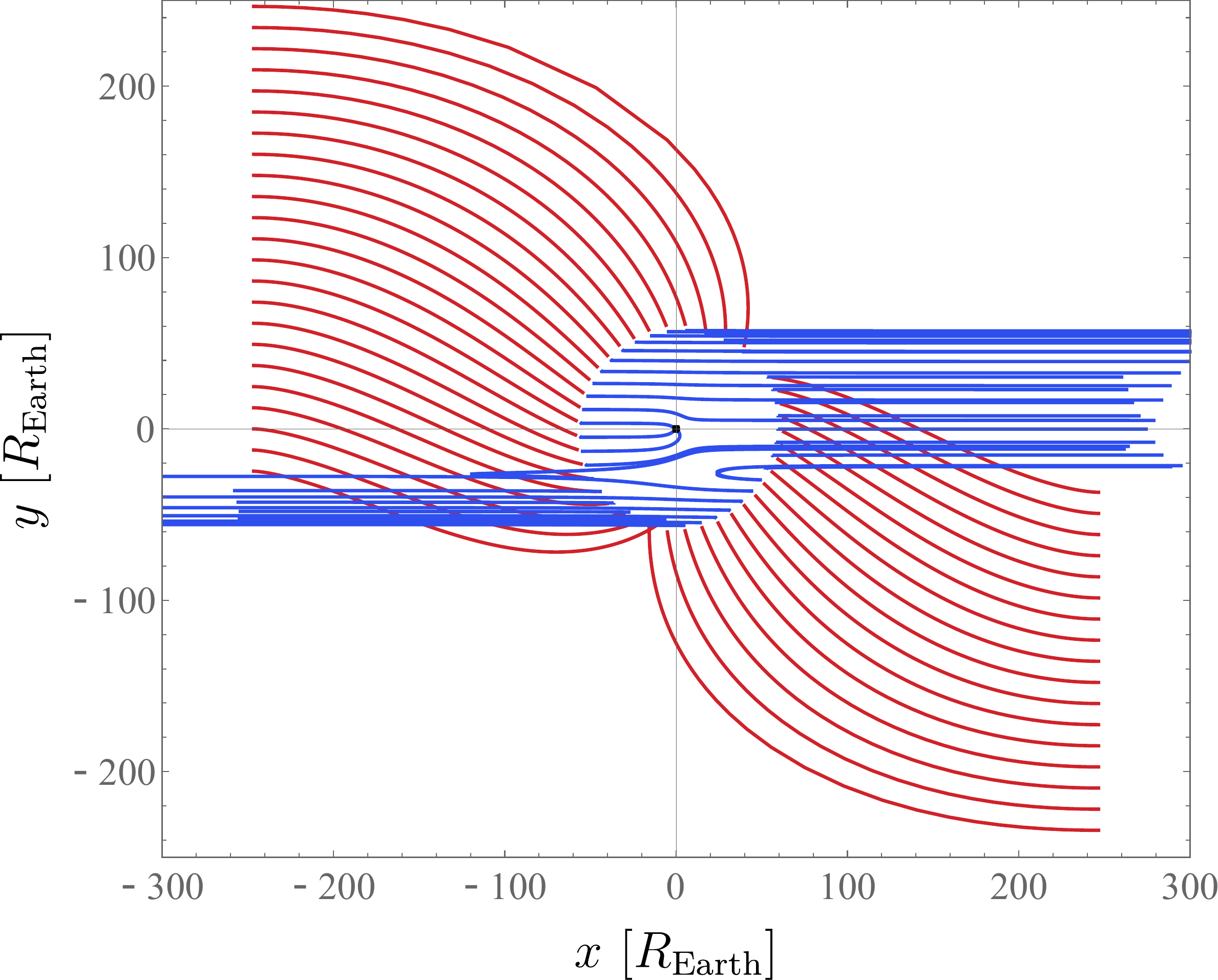}
        \caption{Trajectories of pebbles without wind erosion (top) and with wind erosion (bottom). The planet in the center has Earth's radius ($R_\mathrm{planet}=1 R_\mathrm{Earth}$) and is located at $a=1\,\mathrm{AU}$. The initial pebble radius is $R_\mathrm{pebble}=1\,\mathrm{m}$. The surface energy of the millimeter-sized particles that are constituents of the larger pebble is assumed to be $\gamma=10^{-4}\, \mathrm{N}\,\mathrm{m}^{-1}$. The diameter of the pebbles is shown with the color of the trajectories (red: $2\,\mathrm{m}$, blue: $1\,\mathrm{mm}$). The rapid transition in the case of erosion occurs in the little white gap. For visualization reasons, we chose not to plot the sizes in between.}
        \label{fig:paths}
\end{figure}

We analyzed the outcome of pebble accretion for a wide range of simulation parameters (see Table \ref{tab:simparameter}). After each experimental run, a pebble is categorized as accreted by the planet ($r\leq R_\mathrm{planet}$) or not ($r\geq 1.5 r_\mathrm{init}$). As an example, Fig. \ref{fig:paths} shows the trajectories of $2\,\mathrm{m}$ sized pebbles - or probably better termed more generally "bodies" but we ignore this subtlety here - accreted by an Earth-like planet at $1\,\mathrm{AU}$. The body consists of $1\,\mathrm{mm}$ sized particles. In the top figure wind erosion is deactivated. In the bottom figure wind erosion is enabled. The color of the lines indicates the current diameter of the body (red: $2\,\mathrm{m}$, blue: $1\,\mathrm{mm}$). If wind erosion is deactivated, all bodies in this example are accreted by the planet. By activating the erosion due to gas drag, the bodies are eroded quickly to the minimum size of $1\,\mathrm{mm}$ (see rapid transition between red and blue in Fig. \ref{fig:paths}). With the transition in size - and thus Stokes number $\mathrm{St}=t_\mathrm{c} \Omega_0$ - the bodies change their aerodynamic behavior and couple well to gas. Being deflected by the gas flow, the accretion efficiency is decreased significantly, close to zero in this case.

\begin{figure}
        \centering      \includegraphics[width=0.95\columnwidth]{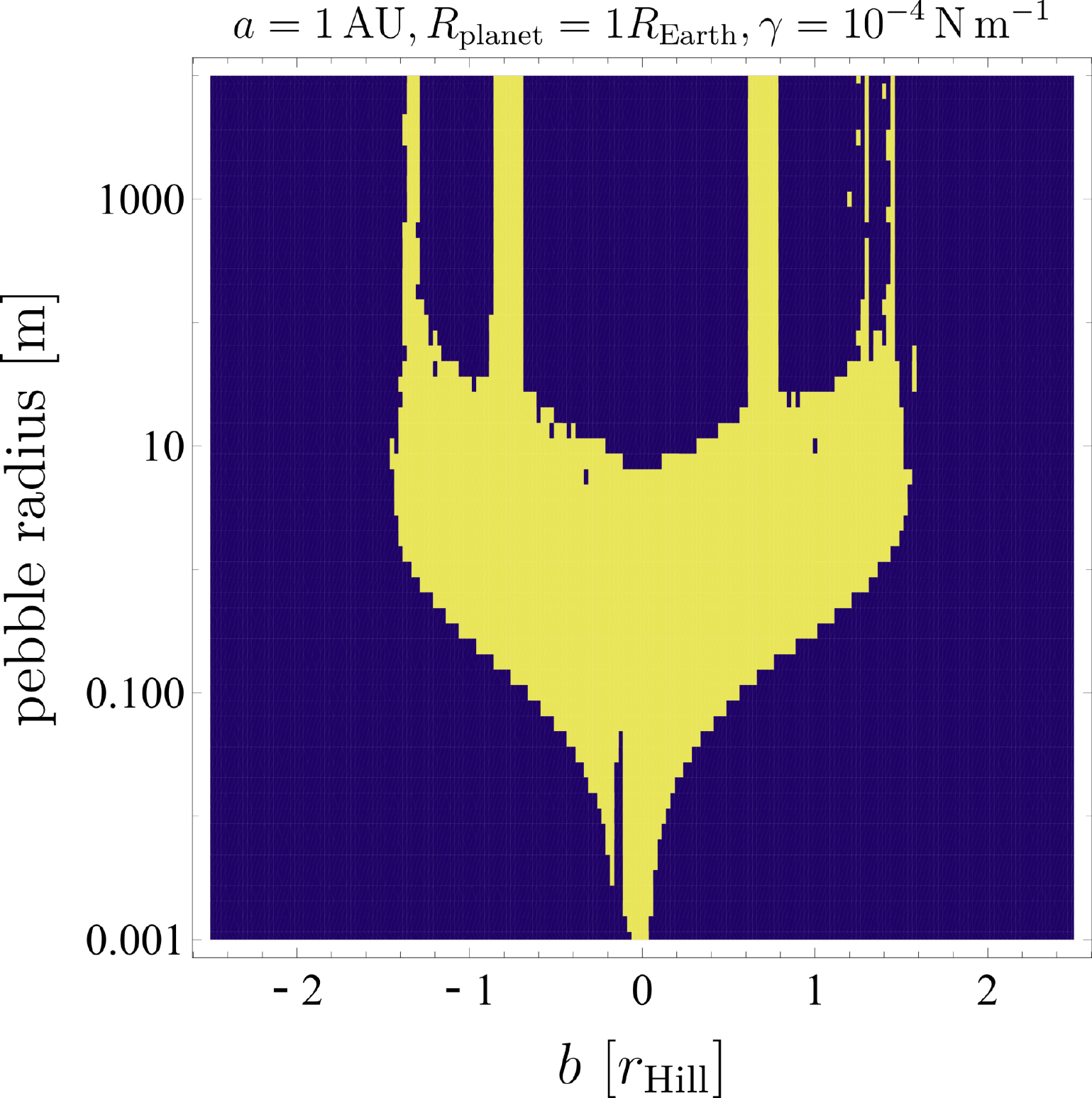}
        \includegraphics[width=0.95\columnwidth]{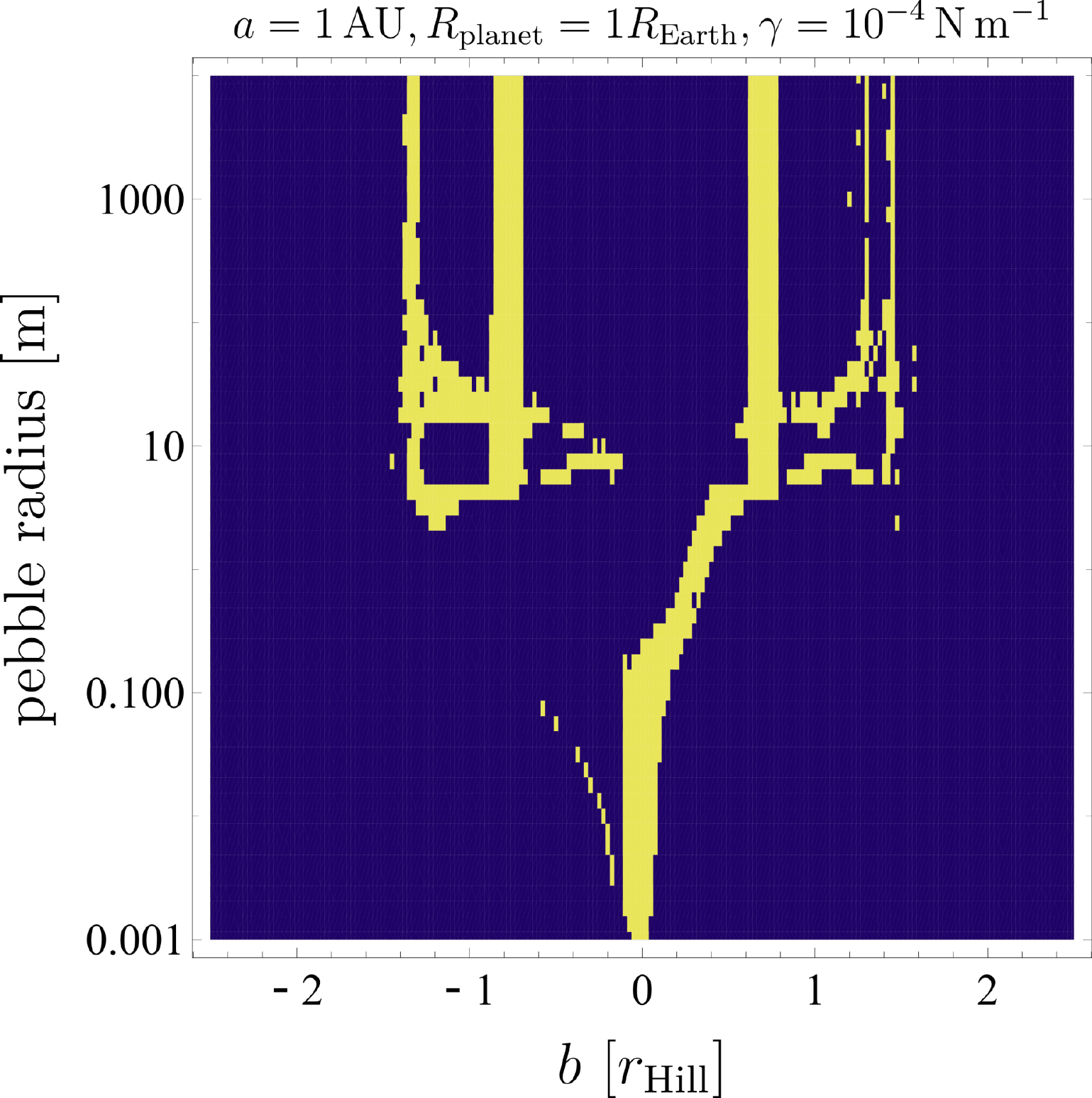}
        \caption{Pebble accretion outcome for a range of initial $y$-coordinates and initial pebble radii. Yellow areas in the plot mark parameters, where pebbles are accreted by the planet, and blue marks parameters where no pebble accretion occurs. The top figure shows the case without wind erosion and the bottom figure the case with erosion. Obviously, wind erosion decreases the parameter range for pebble accretion strongly.}
        \label{fig:parameterplot}
\end{figure}

\begin{figure}
        \centering
        \includegraphics[width=\columnwidth]{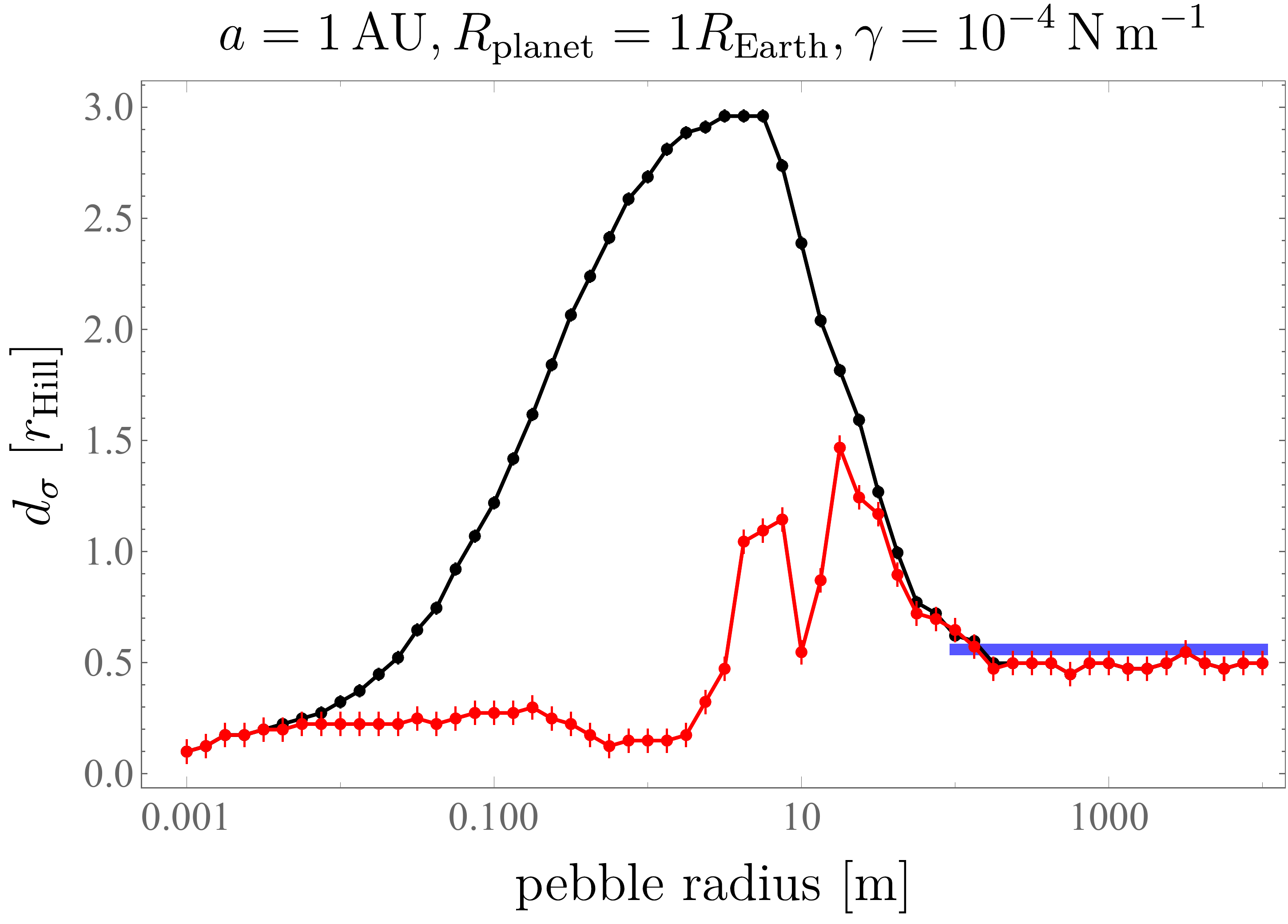}
        \caption{Accretion cross-section diameter $d_\sigma$  dependent on the pebble radius $R_\mathrm{pebble}$ for an Earth-sized planet at $1\,\mathrm{AU}$ with wind erosion (red) and without wind erosion (black). For large pebble radii, both cases converge to an accretion cross-section diameter of about $0.5 r_\mathrm{Hill}$, which is slightly below the theoretical value (blue line) for gravitational focusing \citep{Safronov1972}. We assume that the underestimation is caused by the close encounters that can spiral on highly eccentric orbits onto the planet, but are registered as non-accreting events when they reach a large distance from the planet.}
        \label{fig:crosssectionplot}
\end{figure}

Figure \ref{fig:parameterplot} shows the accretion outcome for an Earth-sized planet at $1\,\mathrm{AU}$ for a wide range of sizes for in-falling bodies ($R_\mathrm{pebble} \in [10^{-3}\,\mathrm{m},10^{4}\,\mathrm{m}]$ and initial $y$-position. The range for the initial $y$-coordinate is kept large enough that all accretion events are considered for the chosen pebble radius range. Here again, the top figure shows the case without wind erosion and the bottom figure the case with wind erosion. There is a clear decrease in pebble accretion if wind erosion is considered. To quantify the influence of wind erosion on pebble accretion, we compare the accretion cross-section diameter $d_\sigma$ as dependent on the pebble radius for the case with erosion ($d_\sigma^\mathrm{e.}$) and without ($d_\sigma^\mathrm{n.e.}$). The cross-section diameter is determined by the width of the yellow area for each pebble radius. An example for a Earth-like planet at $1\,\mathrm{AU}$ is shown in Fig. \ref{fig:crosssectionplot}. 

The black data points represent the case without wind erosion. Small pebbles have a small accretion cross-section, because due to their small coupling times (small Stokes numbers) they mainly follow the gas flow and are not accreted by the planet. The accretion cross-section increases up to a body radius of approximately $10\,\mathrm{m}$ (intermediate Stokes numbers). Pebbles or planetesimals larger than $\sim 10\,\mathrm{m}$ have a rapid decrease in the cross-section size for increasing pebble radius. Objects larger than approximately $100\,\mathrm{m}$ have a constant cross-section diameter. At these large Stokes numbers, they no longer interact with the surrounding gas on relevant timescales. Scattering caused by the planet's gravity dominates. This is consistent with gravitational focusing as indicated by the blue line \citep{Safronov1972}. 

The red data points represent the case with wind erosion. Large objects ($\geq 100\,\mathrm{m}$) are sufficiently stable against wind erosion, so the accretion cross-section diameter does not differ from the case without wind erosion. Small objects ($\leq 1\,\mathrm{cm}$) are coupled well to the gas, so they do not exceed the threshold shear stress for wind erosion. Due to their acceleration on the planet, pebble radii between ($10^{-2}-10^2\,\mathrm{m}$) experience gas-drag-driven erosion. After being eroded to a smaller size, the pebble accretion diameter decreases significantly, due to the aerodynamic change in the pebbles.

\begin{figure}
        \centering      \includegraphics[width=\columnwidth]{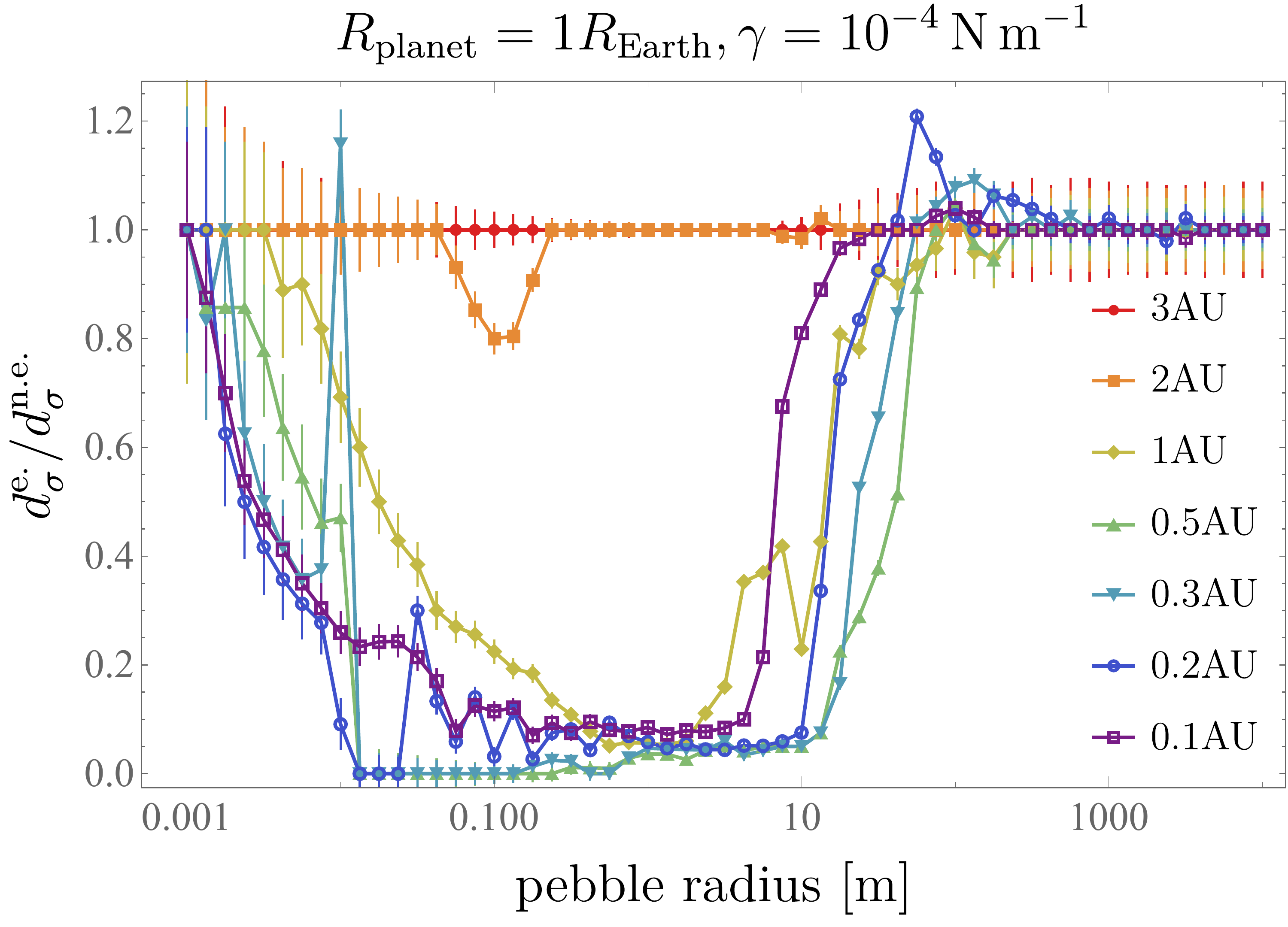}
        \caption{Accretion cross-section diameter $d_\sigma^\mathrm{e.}$ with erosion in relation to $d_\sigma^\mathrm{n.e.}$ without erosion as dependent on the pebble radius $R_\mathrm{pebble}$ for an Earth-sized planet. The semi-major axis $a$ varies between $0.1\,\mathrm{AU}$ (violet) and $3\,\mathrm{AU}$ (red). For planets near to the central star, a significant reduction in accretion efficiency for pebbles with $R_\mathrm{pebble} \lesssim 10\,\mathrm{m}$ can be observed with wind erosion. For a semi-major axis of $a \geq 3\,\mathrm{AU}$ there is no difference between the case with erosion and without.}
        \label{fig:crosssectionfactorplot}
\end{figure}

To illustrate the difference between the case without and with wind erosion, in Fig. 5 we plot the ratio of the accretional cross section diameters  $d_\sigma^\mathrm{e.}$ and $d_\sigma^\mathrm{n.e.}$ as dependent on the pebble radius $R_\mathrm{pebble}$  for an Earth-like planet at different semi-major axes, $a \in [0.1 \, \mathrm{AU}, 3\,\mathrm{AU}]$. For planets near to the central star $a \leq 1\,\mathrm{AU,}$ there is a significant reduction in pebble accretion efficiency for pebble radii between approximately $0.1$ and $10\, \mathrm{m}$ with erosion. Below and above this range the ratio of $d_\sigma^\mathrm{e.}$ to $d_\sigma^\mathrm{n.e.}$ is 1. Wind erosion produces a significant dip in accretion efficiency within this size range. For an increasing semi-major axis, this dip is less distinctive. For $a\geq 3\,\mathrm{AU}$ the dip disappears completely. At these far distances wind erosion does not occur and influence the pebble accretion outcome due to the low gas density.\\

\begin{figure*}
        \centering
        \begin{minipage}[t]{0.32\textwidth}             \includegraphics[width=\textwidth]{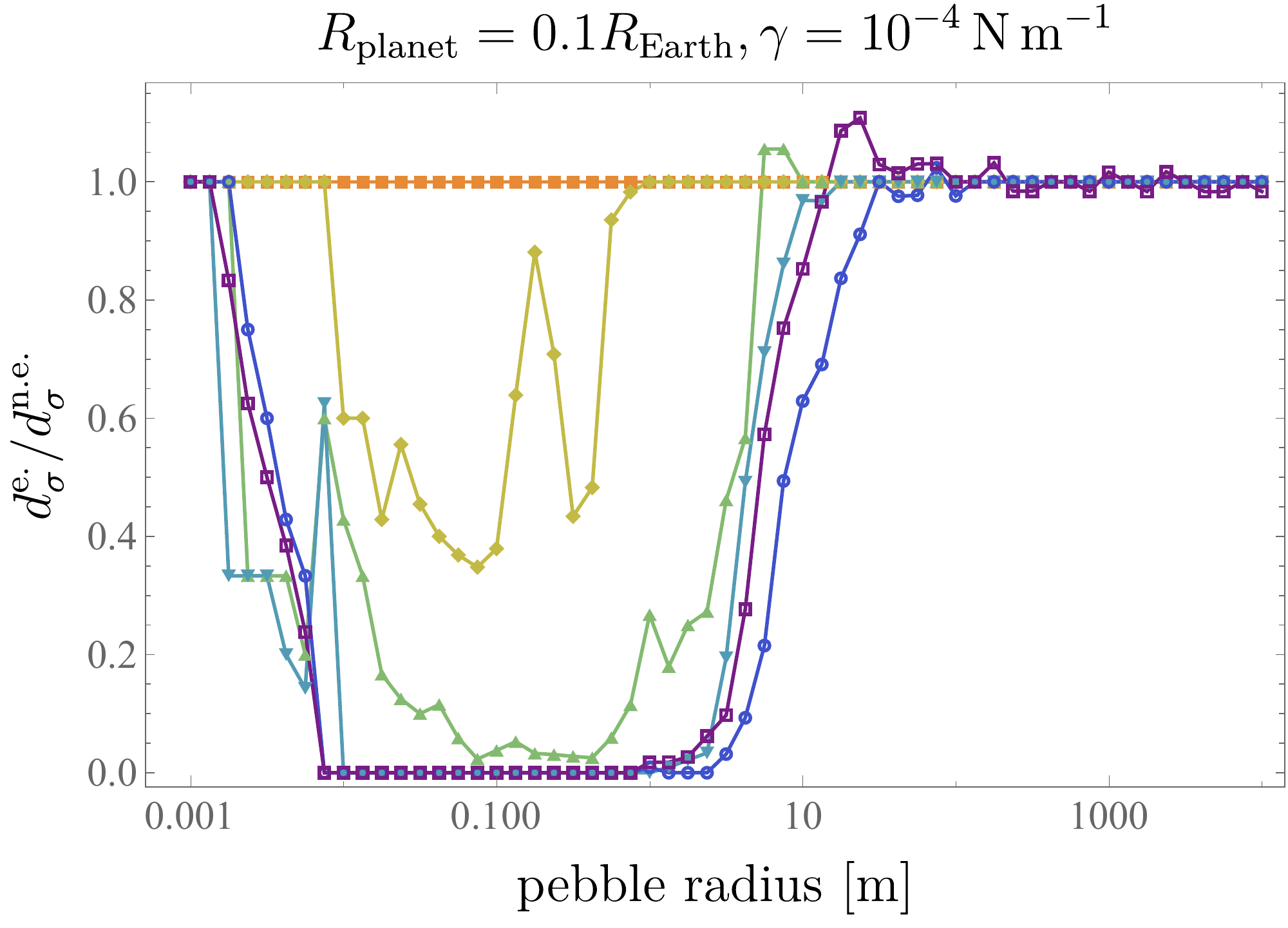}
        \end{minipage}
        \begin{minipage}[t]{0.32\textwidth}             \includegraphics[width=\textwidth]{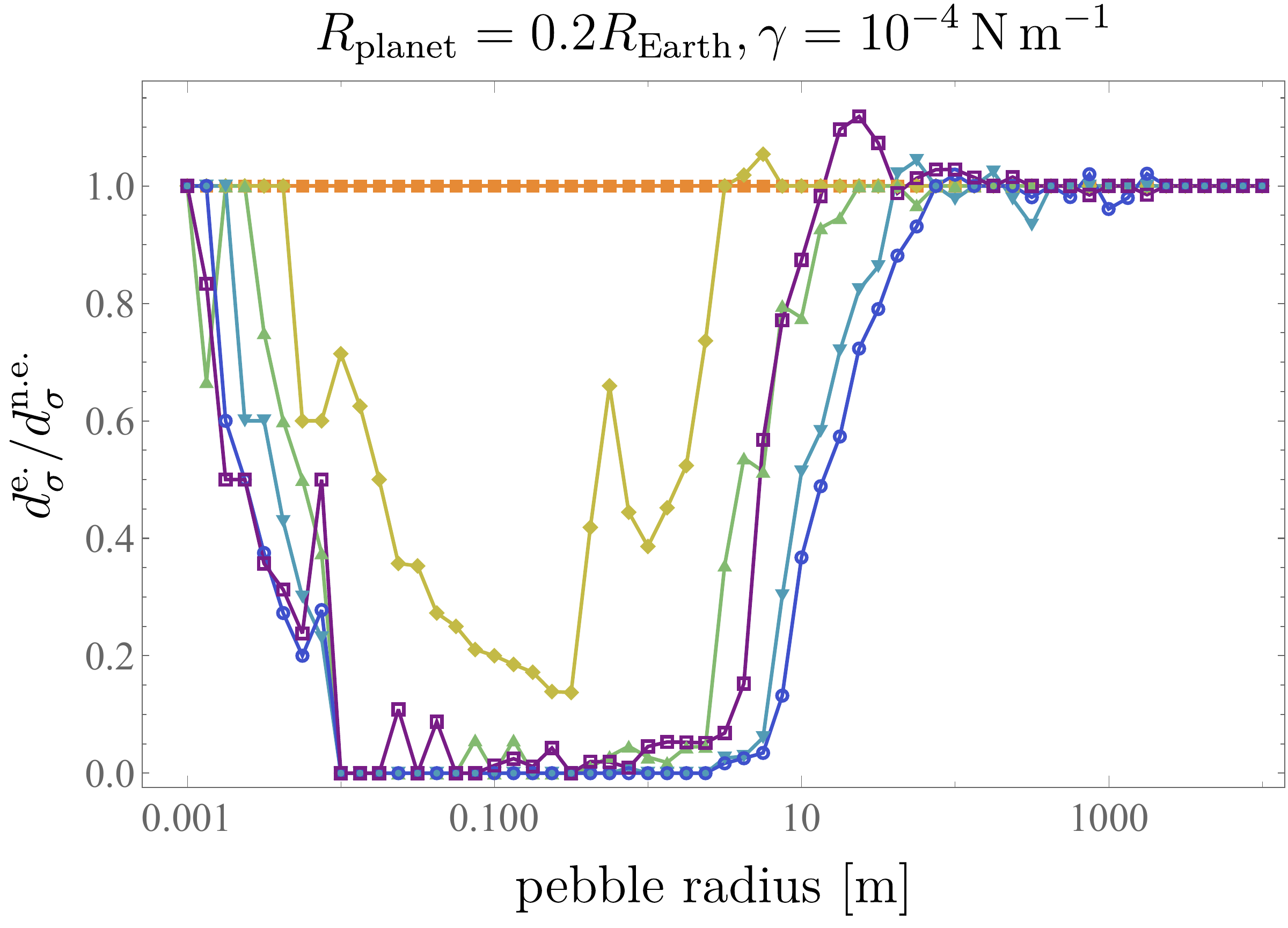}
        \end{minipage}
        \begin{minipage}[t]{0.32\textwidth}             \includegraphics[width=\textwidth]{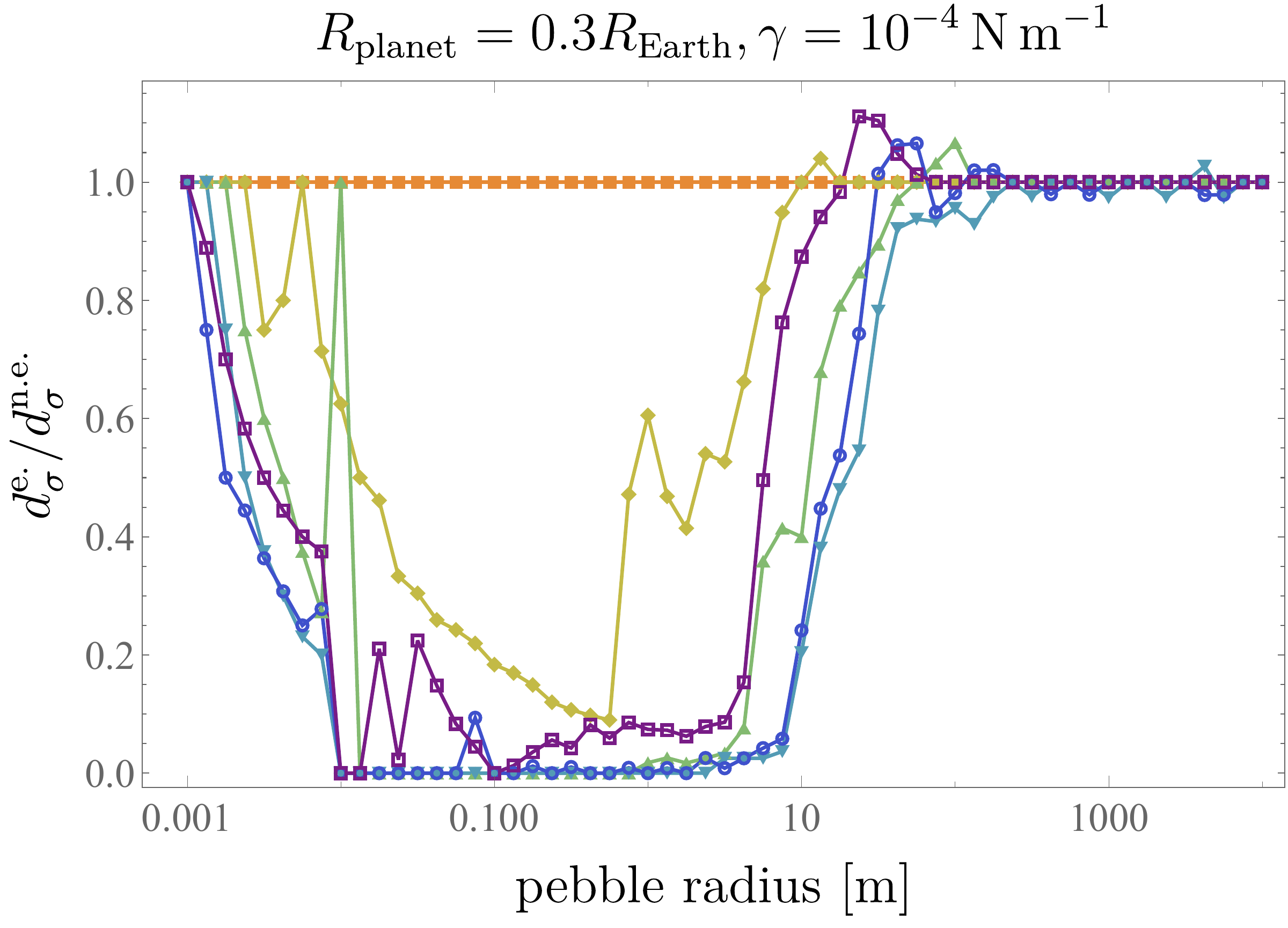}
        \end{minipage}
        
        \begin{minipage}[t]{0.32\textwidth}             \includegraphics[width=\textwidth]{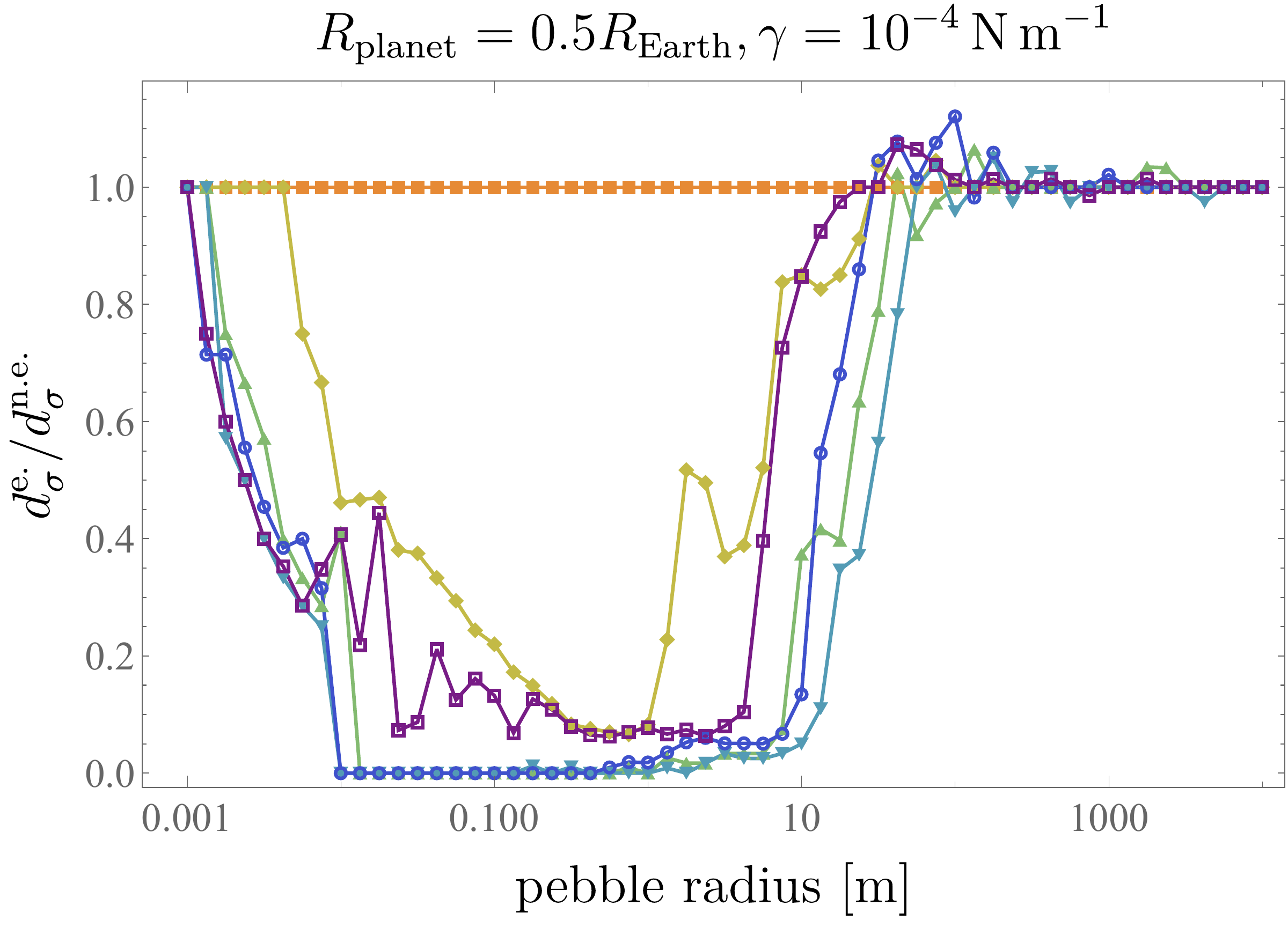}
        \end{minipage}
        \begin{minipage}[t]{0.32\textwidth}             \includegraphics[width=\textwidth]{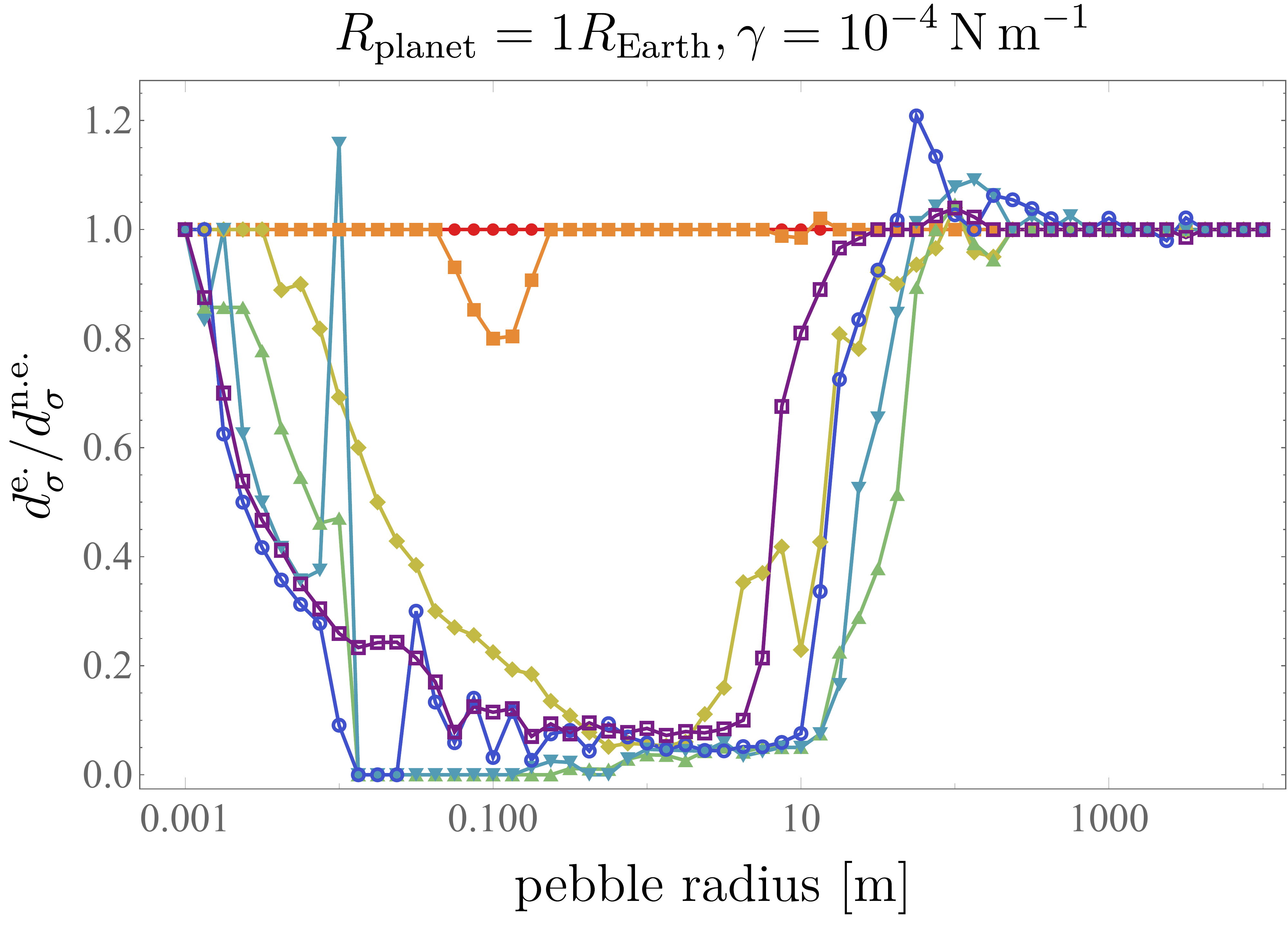}
        \end{minipage}
        \begin{minipage}[t]{0.32\textwidth}             \includegraphics[width=\textwidth]{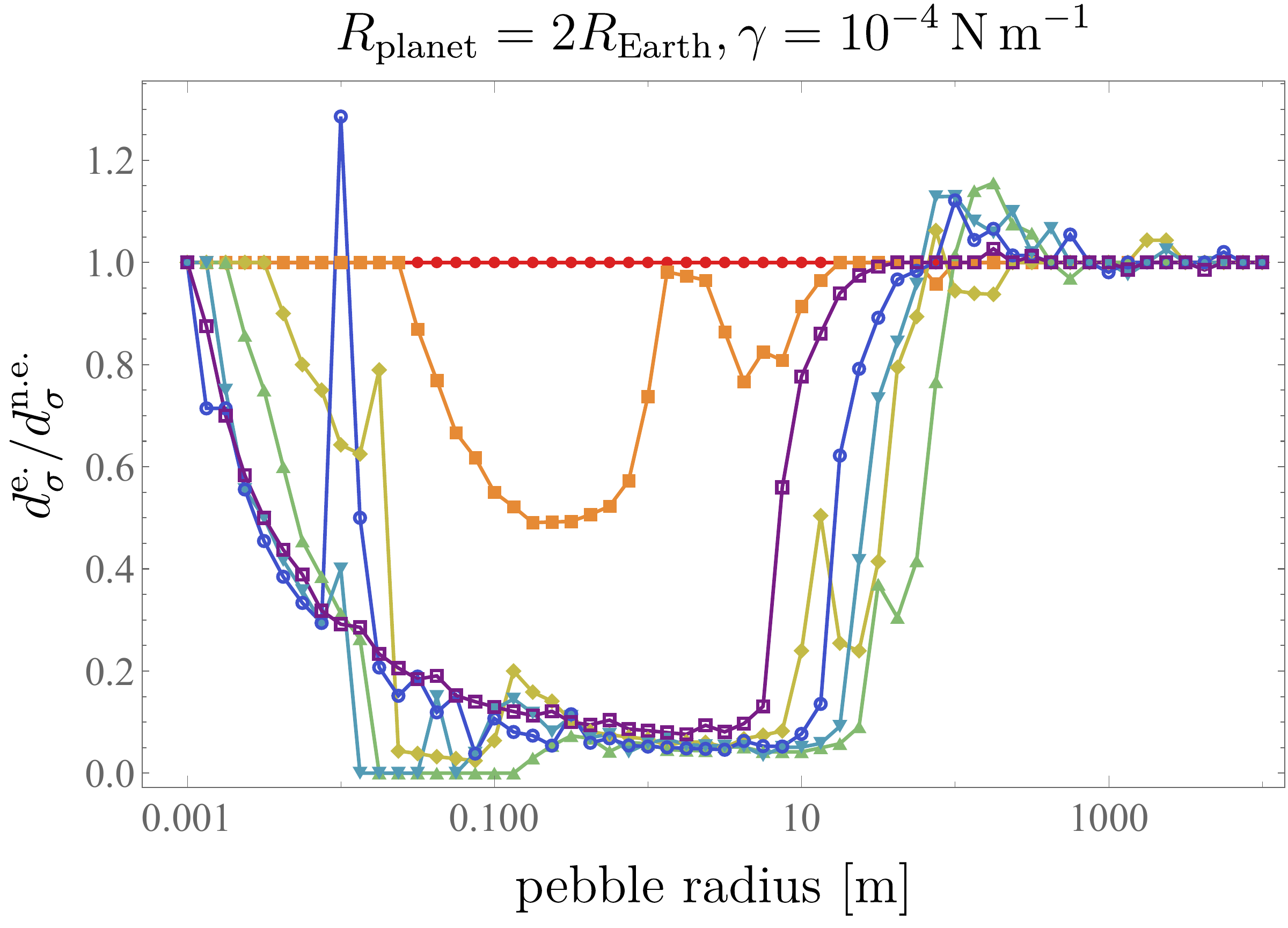}
        \end{minipage}
        
        \begin{minipage}[t]{0.32\textwidth}             \includegraphics[width=\textwidth]{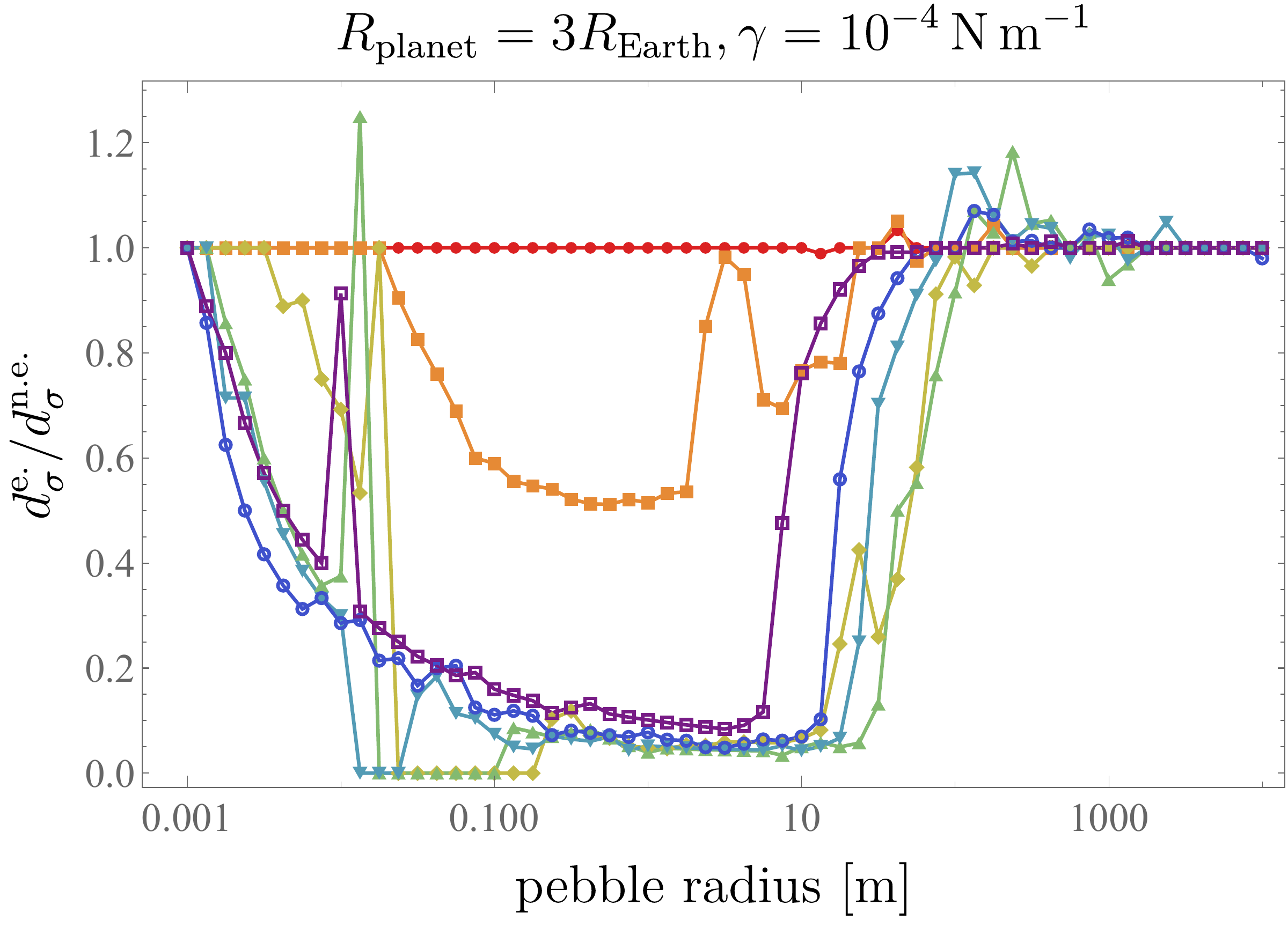}
        \end{minipage}
        \begin{minipage}[t]{0.32\textwidth}             \includegraphics[width=\textwidth]{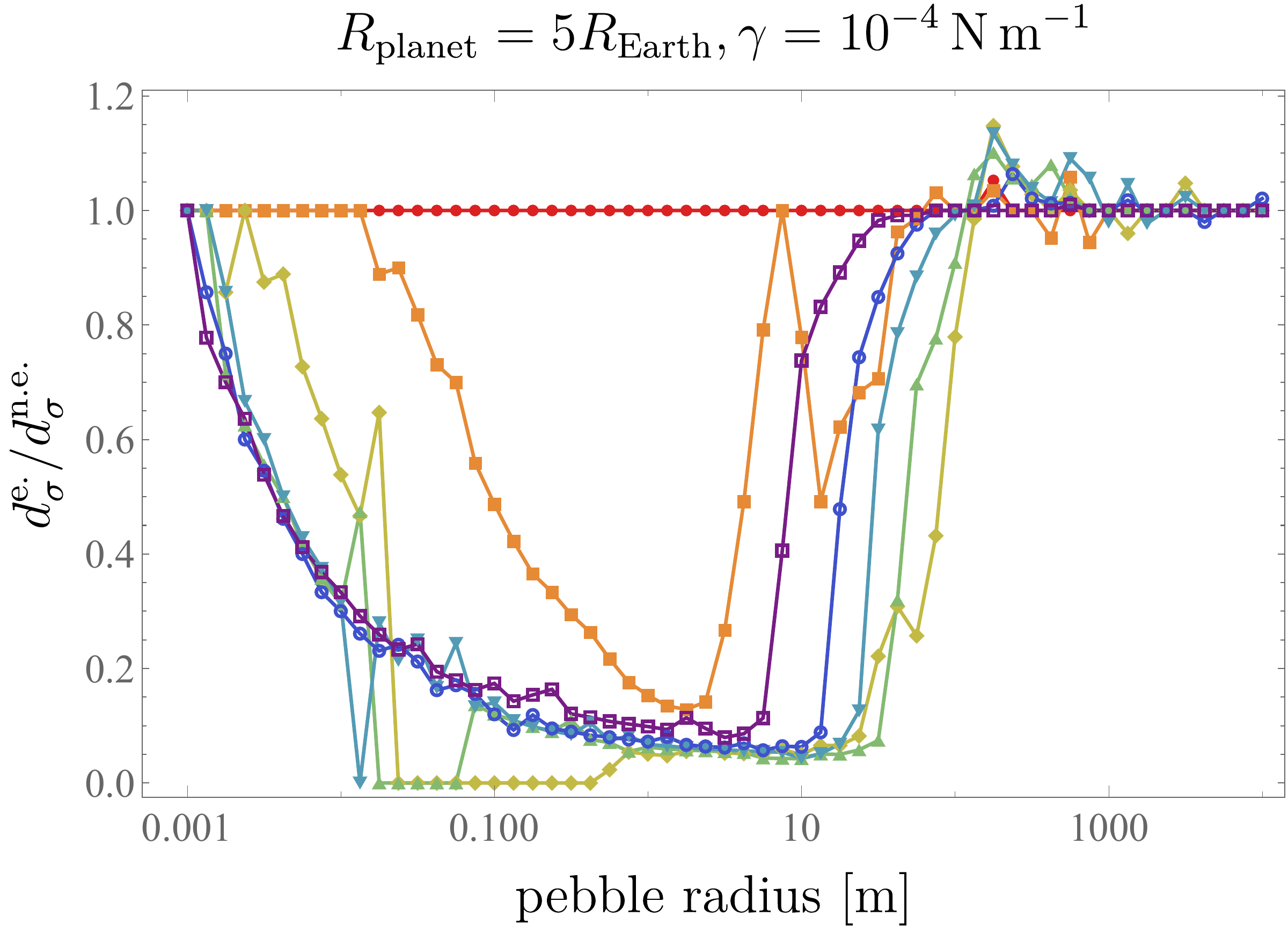}
        \end{minipage}
        \begin{minipage}[t]{0.32\textwidth}             \includegraphics[width=\textwidth]{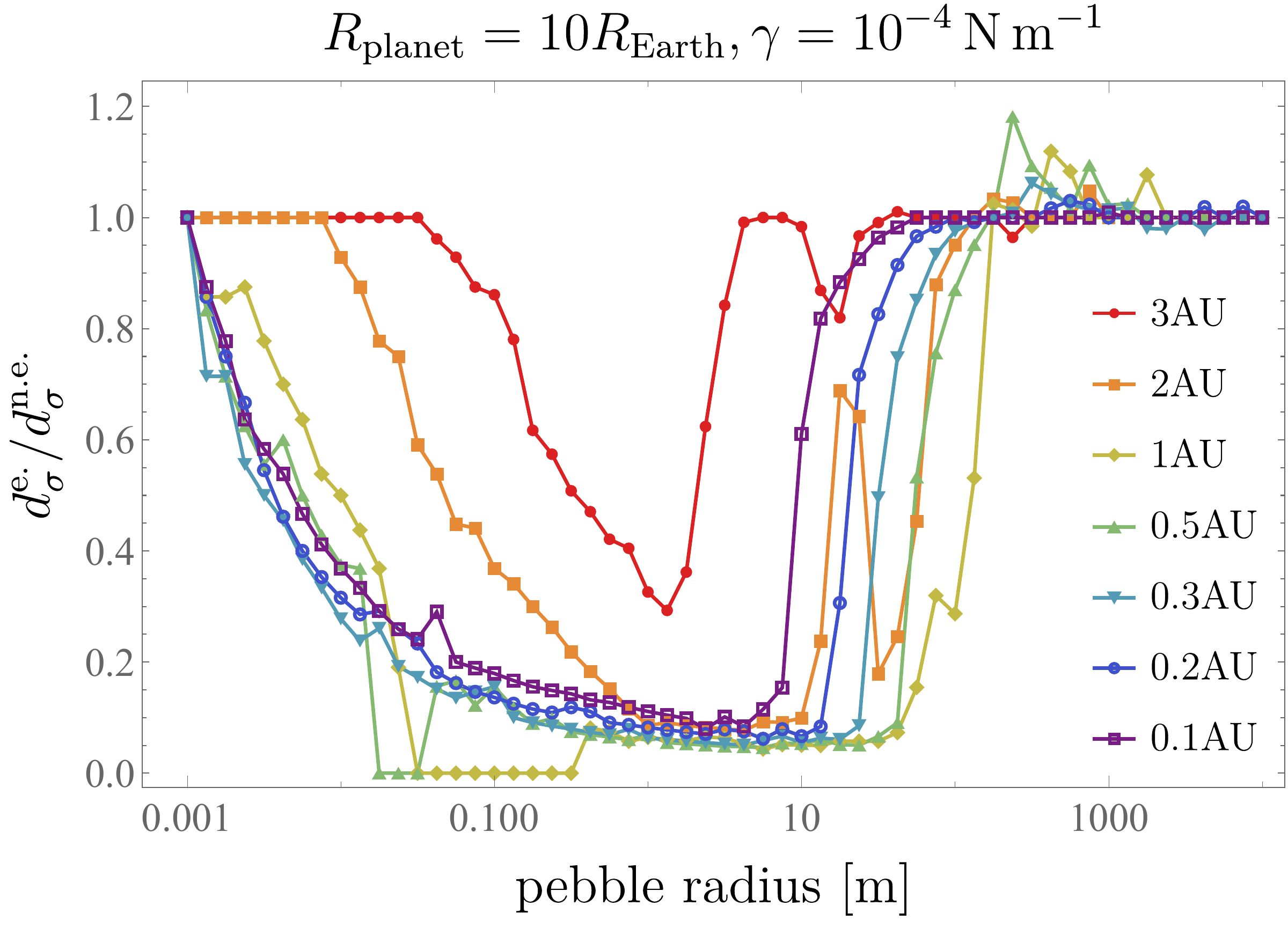}
        \end{minipage}  
        \caption{\label{fig:crossectionfactorplots}Accretion cross-section diameter $d_\sigma^\mathrm{e.}$ with erosion in relation with $d_\sigma^\mathrm{n.e.}$ without erosion depending on the pebble radius $R_\mathrm{pebble}$ for planets with sizes between $0.1$ and $10\,R_\mathrm{Earth}$. The semi-major axis $a$ varies between $0.1\,\mathrm{AU}$ (violet) and $3\,\mathrm{AU}$ (red). The planets do not have an additional planetary atmosphere. For a semi-major axis $a \gtrsim 3\,\mathrm{AU,}$ wind erosion does not affect the pebble accretion outcome (see yellow line at $d_\sigma^\mathrm{e.}/d_\sigma^\mathrm{n.e.}=1$). With increasing planet size $R_\mathrm{planet}$ the pebble radius range, which is affected by the pebble accretion outcome, increases significantly. This can be seen by a shift in the threshold pebble radius around $10\,\mathrm{m}$. Also bigger planets increase the region where wind erosion influences the accretion of pebbles and planetesimals. Error bars (similar to Fig. \ref{fig:crosssectionfactorplot}) are removed for better visualization.}
\end{figure*}

Figure \ref{fig:crossectionfactorplots} shows the ratio $d_\sigma^\mathrm{e.}/d_\sigma^\mathrm{n.e.}$ as dependent on the pebble radius $R_\mathrm{pebble}$ for planets with sizes between $0.1$ and $10\,R_\mathrm{Earth}$. Semi-major axes between $0.1$ and $3\,\mathrm{AU}$ are simulated and no additional planetary atmosphere is considered (Eq. \ref{eq:totalgaspressure} consists only of the first term). For semi-major axes $a \gtrsim 3 \,\mathrm{AU,}$ wind erosion does not affect the pebble accretion outcome. This can be seen by the red data points, which are equal to $d_\sigma^\mathrm{e.}/d_\sigma^\mathrm{n.e.}=1$. The dip, caused by wind erosion, is more pronounced at smaller semi-major axes $a$ and is wider the larger the planet. The threshold pebble radius, which we define as the maximum pebble radius of the wind erosion dip, increases with increasing planet radius. 

\begin{figure*}
        \centering
        \begin{minipage}[t]{0.32\textwidth}             \includegraphics[width=\textwidth]{PA_crosssection_1Earthradius.pdf}
        \end{minipage}
        \begin{minipage}[t]{0.32\textwidth}             \includegraphics[width=\textwidth]{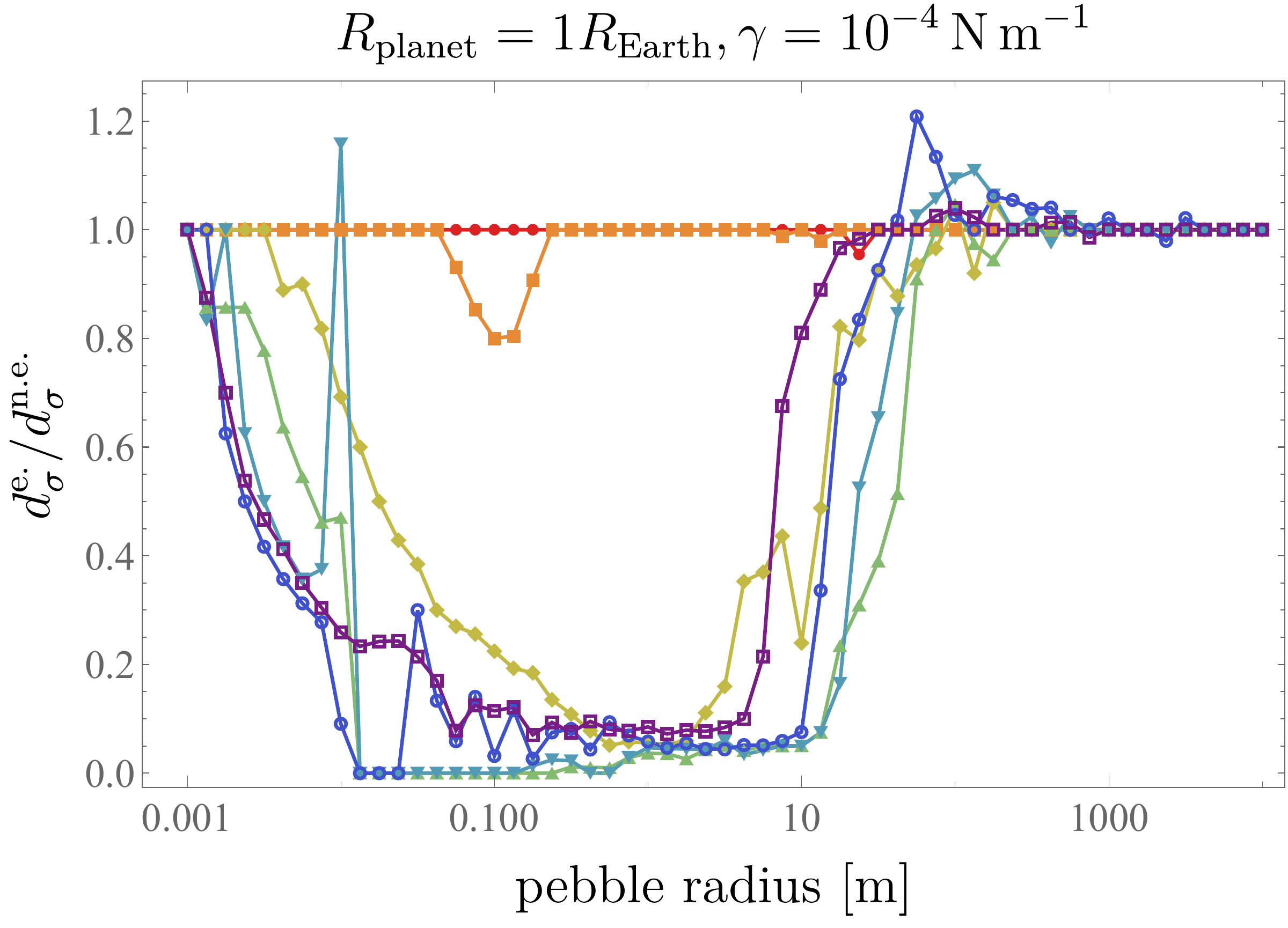}
        \end{minipage}
        \begin{minipage}[t]{0.32\textwidth}             \includegraphics[width=\textwidth]{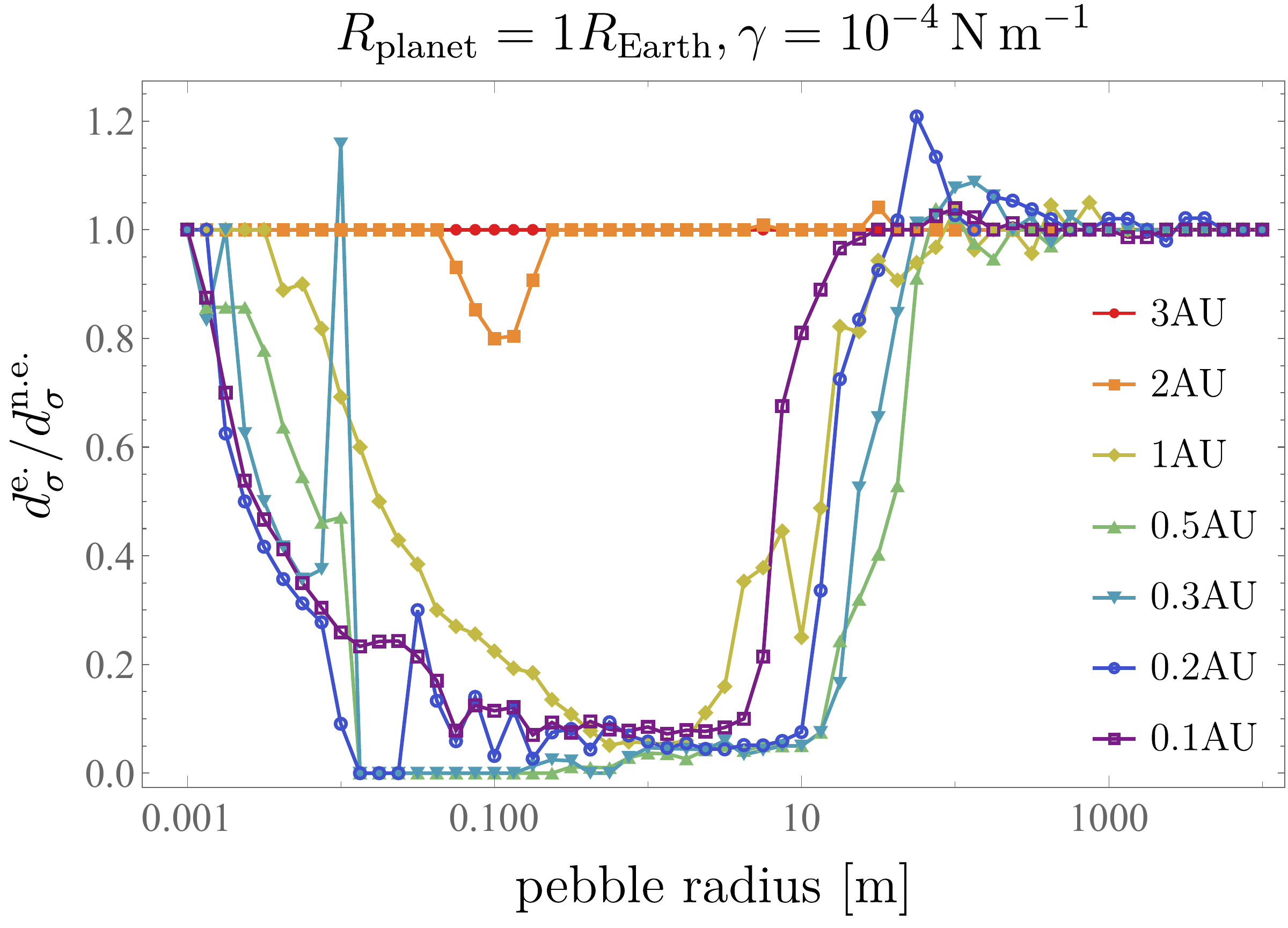}
        \end{minipage}  
        \caption{\label{fig:crossectionfactorplotsatmosphere}Accretion cross-section diameter $d_\sigma^\mathrm{e.}$ with erosion in relation with $d_\sigma^\mathrm{n.e.}$ without erosion depending on the pebble radius $R_\mathrm{pebble}$ for an Earth-sized planet with no atmosphere (left), with Martian atmosphere ($10^3\,\mathrm{Pa}$; middle), and with Earth's atmosphere ($10^5\,\mathrm{Pa}$; right). The accretion outcome of these three cases does not differ significantly. This indicates that wind erosion of pebbles and planetesimals, accreted by the planet, occurs at distances far away from the planet where the planetary atmosphere has no additional influence on the local gas pressure (see Eq. \ref{eq:totalgaspressure}). The semi-major axis $a$ varies between $0.1\,\mathrm{AU}$ (violet) and $3\,\mathrm{AU}$ (red). Error bars (similar to Fig. \ref{fig:crosssectionfactorplot}) are removed for better visualization.}
\end{figure*}

The data presented so far have not taken the additional planetary atmosphere into account. We consider that an Earth-sized planet might hang on to an additional planetary atmosphere. Therefore, we carried out two simulation runs with $10^3\,\mathrm{Pa}$ and $10^5\,\mathrm{Pa}$ gas pressure at the planet's surface and compared the results with that of the case without planetary atmosphere. The thin atmosphere is similar to the Martian atmosphere while the thicker one is comparable with Earth's atmosphere. Figure \ref{fig:crossectionfactorplotsatmosphere} compares the results for the different planetary atmospheres (left: no atmosphere, middle: $10^3\,\mathrm{Pa}$, right: $10^5\,\mathrm{Pa}$). For the probed parameter space ($R_\mathrm{pebble} \in [10^{-3}\,\mathrm{m},10^{4}\,\mathrm{m}]$ and $a \in [0.1\,\mathrm{AU},3\,\mathrm{AU}]$), we see no significant difference in the pebble accretion outcome. This indicates that wind erosion of pebbles and planetesimals, which are accreted by the planet, occurs at distances far away from the planet where the planetary atmosphere has no additional influence on the local gas pressure (see Eq. \ref{eq:totalgaspressure}). This means that for future pebble accretion simulations the local gas density in the protoplanetary disk is more important than the atmosphere of the planet. The reason for this is that the planetary atmosphere only becomes relevant at very small distances from the planetary surface, so that pebbles and planetesimals eroded at these distances are accreted anyway.

\section{Conclusion}

We studied the influence of gas-drag-driven wind erosion on pebble and planetesimal accretion in numerical simulations. For bodies consisting of millimeter-sized particles, wind erosion can be a destructive process, which can lead to complete dissolution into the individual parts. During pebble or planetesimal accretion, where the pebble or the planetesimal can reach high relative velocities to the gas due to the gravitational acceleration of the planet, wind erosion is a major driver for the accretion outcome. For a semi-major axis  of $a<3\,\mathrm{AU}$, we observe that wind erosion decreases the accretion efficiency for pebbles smaller $\lesssim 10 \, \mathrm{m}$ significantly. We also observe that it is not important for the accretion outcome whether the planet has an additional planetary atmosphere or not, at least the kind of atmosphere the Earth has. The wind erosion dip in the accretion efficiency is characterized by the threshold pebble radius and this quantity is dependent on the planetary radius $R_\mathrm{planet}$ and semi-major axis $a$.

\begin{acknowledgements}
This project is funded by DLR space administration with funds provided by the BMWi under grant 50 WM 1760. We thank Remo Burn for a constructive discussion of these processes. We appreciate the constructive review by the anonymous referee.
\end{acknowledgements}

\bibliographystyle{aa}
\bibliography{demirci}

\begin{thebibliography}{31}
\expandafter\ifx\csname natexlab\endcsname\relax\def\natexlab#1{#1}\fi

\bibitem[{{Alibert} {et~al.}(2018){Alibert}, {Venturini}, {Helled}, {Ataiee},
  {Burn}, {Senecal}, {Benz}, {Mayer}, {Mordasini}, {Quanz}, \&
  {Sch{\"o}nb{\"a}chler}}]{Alibert2018}
{Alibert}, Y., {Venturini}, J., {Helled}, R., {et~al.} 2018, Nature Astronomy,
  2, 873

\bibitem[{Bitsch {et~al.}(2019)Bitsch, Izidoro, Johansen, Raymond, Morbidelli,
  Lambrechts, \& Jacobson}]{Bitsch2019}
Bitsch, B., Izidoro, A., Johansen, A., {et~al.} 2019, A\&A, 623, A88

\bibitem[{Bitsch {et~al.}(2018)Bitsch, Lambrechts, \& Johansen}]{Bitsch2018}
Bitsch, B., Lambrechts, M., \& Johansen, A. 2018, A\&A, 609, C2

\bibitem[{Brown \& Lawler(2003)}]{Brown2003}
Brown, P.~P. \& Lawler, D.~F. 2003, Journal of Environmental Engineering, 129,
  222

\bibitem[{{Br{\"u}gger} {et~al.}(2020){Br{\"u}gger}, {Burn}, {Coleman},
  {Alibert}, \& {Benz}}]{Bruegger2020}
{Br{\"u}gger}, N., {Burn}, R., {Coleman}, G., {Alibert}, Y., \& {Benz}, W.
  2020, arXiv e-prints, arXiv:2006.04121

\bibitem[{Davies(1945)}]{Davies1945}
Davies, C.~N. 1945, Proceedings of the Physical Society, 57, 259

\bibitem[{{Demirci} {et~al.}(2019){Demirci}, {Kruss}, {Teiser}, {Bogdan},
  {Jungmann}, {Schneider}, \& {Wurm}}]{Demirci2019}
{Demirci}, T., {Kruss}, M., {Teiser}, J., {et~al.} 2019, MNRAS, 484, 2779

\bibitem[{Demirci {et~al.}(2020)Demirci, Schneider, Steinpilz, Bogdan, Teiser,
  \& Wurm}]{Demirci2020}
Demirci, T., Schneider, N., Steinpilz, T., {et~al.} 2020, MNRAS, 493, 5456

\bibitem[{Demirci {et~al.}(2017)Demirci, Teiser, Steinpilz, Landers, Salamon,
  Wende, \& Wurm}]{Demirci2017}
Demirci, T., Teiser, J., Steinpilz, T., {et~al.} 2017, The Astrophysical
  Journal, 846, 48

\bibitem[{Hayashi(1981)}]{Hayashi1981}
Hayashi, C. 1981, Progress of Theoretical Physics Supplement, 70, 35

\bibitem[{Johansen \& Lambrechts(2017)}]{Johansen2017}
Johansen, A. \& Lambrechts, M. 2017, Annual Review of Earth and Planetary
  Sciences, 45, 359

\bibitem[{Johansen {et~al.}(2015)Johansen, Low, Lacerda, \&
  Bizzarro}]{Johansen2015}
Johansen, A., Low, M.-M.~M., Lacerda, P., \& Bizzarro, M. 2015, Science
  Advances, 1

\bibitem[{Johnson {et~al.}(1971)Johnson, Kendall, Roberts, \&
  Tabor}]{Johnson1971}
Johnson, K.~L., Kendall, K., Roberts, A.~D., \& Tabor, D. 1971, Proceedings of
  the Royal Society of London. A. Mathematical and Physical Sciences, 324, 301

\bibitem[{Kruss {et~al.}(2016)Kruss, Demirci, Koester, Kelling, \&
  Wurm}]{Kruss2016}
Kruss, M., Demirci, T., Koester, M., Kelling, T., \& Wurm, G. 2016, ApJ, 827,
  110

\bibitem[{Kruss {et~al.}(2017)Kruss, Teiser, \& Wurm}]{Kruss2017}
Kruss, M., Teiser, J., \& Wurm, G. 2017, A\&A, 600, A103

\bibitem[{Lambrechts \& Johansen(2012)}]{Lambrechts2012}
Lambrechts, M. \& Johansen, A. 2012, A\&A, 544, A32

\bibitem[{Liu {et~al.}(2019)Liu, Ormel, \& Johansen}]{Liu2019}
Liu, B., Ormel, C.~W., \& Johansen, A. 2019, A\&A, 624, A114

\bibitem[{Ormel \& Klahr(2010)}]{Ormel2010}
Ormel, C.~W. \& Klahr, H.~H. 2010, A\&A, 520, A43

\bibitem[{Safronov(1972)}]{Safronov1972}
Safronov, V. 1972, Evolution of the protoplanetary cloud and formation of the
  earth and the planets, NASA technical translation (Israel Program for
  Scientific Translations)

\bibitem[{Schaffer {et~al.}(2020)Schaffer, Johansen, Cedenblad, Mehling, \&
  Mitra}]{Schaffer2020}
Schaffer, N., Johansen, A., Cedenblad, L., Mehling, B., \& Mitra, D. 2020,
  Erosion of planetesimals by gas flow

\bibitem[{Schlichting \& Gersten(2006)}]{Schlichting2006}
Schlichting, H. \& Gersten, K. 2006, Grenzschicht-Theorie (Springer)

\bibitem[{Schneider {et~al.}(2019)Schneider, Wurm, Teiser, Klahr, \&
  Carpenter}]{Schneider2019}
Schneider, N., Wurm, G., Teiser, J., Klahr, H., \& Carpenter, V. 2019, ApJ,
  872, 3

\bibitem[{Schreiber \& Klahr(2018)}]{Schreiber2018}
Schreiber, A. \& Klahr, H. 2018, ApJ, 861, 47

\bibitem[{Simon {et~al.}(2016)Simon, Armitage, Li, \& Youdin}]{Simon2016}
Simon, J.~B., Armitage, P.~J., Li, R., \& Youdin, A.~N. 2016, ApJ, 822, 55

\bibitem[{Skorov \& Blum(2012)}]{Skorov2012}
Skorov, Y. \& Blum, J. 2012, Icarus, 221, 1

\bibitem[{Squire \& Hopkins(2018)}]{Squire2018}
Squire, J. \& Hopkins, P.~F. 2018, MNRAS, 477, 5011

\bibitem[{{Valletta} \& {Helled}(2019)}]{Valletta2019}
{Valletta}, C. \& {Helled}, R. 2019, ApJ, 871, 127

\bibitem[{{Venturini} \& {Helled}(2020)}]{Venturini2020}
{Venturini}, J. \& {Helled}, R. 2020, \aap, 634, A31

\bibitem[{Weidenschilling(1977)}]{Weidenschilling1977}
Weidenschilling, S.~J. 1977, MNRAS, 180, 57

\bibitem[{Yang {et~al.}(2017)Yang, Johansen, \& Carrera}]{Yang2017}
Yang, C.-C., Johansen, A., \& Carrera, D. 2017, A\&A, 606, A80

\bibitem[{Youdin \& Goodman(2005)}]{Youdin2005}
Youdin, A.~N. \& Goodman, J. 2005, ApJ, 620, 459

\end{thebibliography}

\end{document}